\documentclass[12pt,doublespace]{article}
\begin{document}
\begin{center}
PARAMETER MISMATCHES, CHAOS SYNCHRONIZATION AND FAST DYNAMIC LOGIC GATES \\
E.M.Shahverdiev $^{1,2,*}$ \\
$^{1}$School of Electronic Engineering,Bangor University, Dean St.,Bangor, LL57 1UT, Wales, UK\\
$^{2}$Institute of Physics, H.Javid Avenue,33, Baku, AZ1143, Azerbaijan\\
$^{*}$e.shahverdiev@bangor.ac.uk\\
~\\
ABSTRACT
\end{center}
By using chaos synchronization between non-identical multiple time delay semiconductor lasers with optoelectronic feedbacks, we demonstrate numerically how fast dynamic logic gates can be constructed. The results may be helpful to obtain a computational hardware with reconfigurable properties.\\
~\\
Key words:Chaos synchronization;parameter mismatches; semiconductor lasers; dynamic logic gates, multiple variable time-delays; optoelectronic feedbacks.\\
~\\
PACS number(s):42.55.Px, 42.65.Sf, 05.45.Xt, 42.60.Mi,05.45.Gg, 05.45.Vx\\
\begin{center}
I. INTRODUCTION
\end{center}
\indent Recently due to its fundamental and applied interests chaos synchronization, as one of successful chaos control methods has become a central topic in the nonlinear dynamics [1-2]. Examples of sychronization including chaos synchronization can be found in the activity of the pacemaker heart cells, the circadian rhythms, laser arrays, Josephson junction arrays, etc. [2]. Chaos synchronization can also be applied to improve and optimize the performance of the nonlinear systems [2].\\
Also, lately chaos-based communication systems is emerging as an alternative technique to improve security in such systems, especially after the recent field demonstration using a metropolitan fibre network [3]. For message decoding in such schemes one has to be able to synchronize the transmitter and receiver lasers [2-4].\\
\indent Quite recent new direction for the application of chaos synchronization is to harness chaos for flexible chaos computing. That is to use chaos synchronization to obtain reconfigurable dynamic logic gates, which could make chaos synchronization-based computers more flexible than statically wired hardware systems, see, e.g. [5] and references there-in.\\
\indent In recent work [5] the scheme based on chaos synchronization in Chua circuits is proposed for the implementation of dynamical and flexible logic gates, namely for the fundamental NOR logic gate (from which all gates can be constructed) and for other logic gates such as XOR, AND, NAND, and OR. However, it is clear that the operational speeds achievable by such Chua circuit based computers will not be high enough to present wide practical interest. Although we emphasize that recently fast computing systems based on chaos synchronization is reported with optoelectronic realization of NOR logic gate using chaotic two-section lasers [6]. Namely in [6] theoretical construction of the fundamental NOR gate using laser diodes subject to optical injection is reported. Two self-pulsating laser diodes are commonly driven by a monochromatic light beam resulting in chaos and the NOR logic gate is implemented by appropriately synchronizing the two chaotic attractors.\\
\indent However, as emphasized above the operational speed of the scheme considered in [5] is slow, and in [6] only the fundamental logic NOR is considered, and flexible dynamical construction of the other logic gates were not investigated. As emphasized in the concluding parts of [5]''{\it Clearly, the operational 
speeds attainable by such a dynamic logic cell will be determined by the natural time scales of the dynamical system. So it is evident the scheme could potentially work at considerable speeds if one can implement it on fast dynamical systems, such as electronic circuits operating in the GHz regime and the semiconductor or fiber lasers yielding chaotic subnanosecond or picosecond pulses}''.\\
\indent It is also noted that the systems considered in both [5] and [6] are based on the ordinary differential equations (ODE). Chaos in such systems, although discovered first is not as complex and rich as in time-delayed systems. Time-delayed systems which arise from a realistic consideration of finite communication times are the more adequate models of the interacting systems. The complexity of such systems can also be much more higher than that of the ODE systems. This property makes the time-delayed systems more attractive in application, especially in chaos-based communication systems. As demonstrated in [7], time delay systems (including variable delays) with multiple feedbacks are even more promising from the security point of view chaos based communication systems. In addition, in multiple time delayed systems with the availability of more parameters it is possible to change several parameters in parallel for multi-bit logic synthesis.\\
\indent In this work we explore the possibility of construction of the flexible dynamics logic gates using synchronization between non-identical multiple time delay lasers with optoelectronic feedbacks with natural.Optoelectronic feedback lasers are the dynamical systems with natural time scales of the order of nanoseconds. Here we exploit complete synchronization to construct flexible dynamic logic gates.\\
In [5] studying the synchronization error dynamics is the key part in building logic gate(s). Here we also exploit 
the cross-correlation function [8] for the different level of parameter mismatches between the laser systems to be synchronized to synthesize the flexible logic gates. Table I gives the output corresponding to the two inputs $L_{1}$ and $L_{2}$ for NOR, AND, and XOR logic. We remind that NOR is the fundamental logic gate, from which all gates may be constructed; the logic gates AND and XOR yield the building block of arithmetic processing. We also note that the four distinct possible input sets (0,0),(0,1),(1,0) and (1,1) reduce to three conditions as (0,1) and (1,0) are symmetric.\\
\begin{center}
II. SYSTEM MODEL
\end{center}
\indent Consideration is given to a system composed by two coupled identical single mode semiconductor lasers subject to optoelectronic feedbacks. In a master/slave configuration the optical power emitted by each laser is divided into several parts, detected, amplified, and added to their own injection current; also the optical power emitted by the master laser is detected, amplified, and added to the bias current of the slave laser (figure 1) Thus the dynamics of the double time delay master/transmitter laser is governed by the following systems [9]
\begin{equation}
\frac{dS_{1}}{dt}=(\Gamma g_{1} -\gamma_{c})S_{1}
\end{equation}
\begin{equation}
\frac{dN_{1}}{dt}=I_{1} - \gamma_{s1} N_{1} - g_{1}S_{1} + \gamma_{c}(k_{1} S_{1}(t-\tau_{1}) + k_{2} S_{1}(t-\tau_{2}))
\end{equation}
The slave/receiver laser is described by  
\begin{equation}
\frac{dS_{2}}{dt}=(\Gamma g_{2} -\gamma_{c})S_{2}
\end{equation}
\begin{equation}
\frac{dN_{2}}{dt}=I_{2} - \gamma_{s2} N_{2} - g_{2}S_{2} + \gamma_{c}(k_{3} S_{2}(t-\tau_{1}) + k_{4} S_{2}(t-\tau_{2}) + KS_{1}(t-\tau_{3})) 
\end{equation}
where subindicies 1,2 distinguish between the transmitter and receiver; $S_{1,2}$ is the photon density; $N_{1,2}$ is the carrier density;
$g_{1,2}$ is the material gain;$\gamma_{c}$ is the cavity decay rate;$\gamma_{s}$ is the carrier relaxation rate;$\Gamma$ is the confinement factor of the laser waveguide. $I_{1,2}$ is the bias current (in units of the electron charge);
$k_{1,2}$ and $k_{3,4}$ are the feedback rates for the transmitter and receiver systems, respectively; $\tau_{1,2}$
are the feedback delay times in the transmitter and receiver systems; $K$ is the coupling rate between the
transmitter and the receiver.$\tau_{3}$ is the time of flight between lasers. In a wide operation range the material gain $g$ can be expanded as 
\begin{equation}
g\approx g_{0} + g_{n}(N-N_{0}) + g_{p}(S-S_{0}),
\end{equation}
where $g_{0}=\gamma_{c}/\Gamma $ is the material gain at the solitary threshold; $g_{n}=\partial g/\partial N >0$ is the differential gain parameter; 
$g_{p}=\partial g/\partial S <0$ is the nonlinear gain parameter; $N_{0}$ is the carrier density at threshold; $S_{0}$ is the free-running intra-cavity photon density when the lasers are decoupled; the parameters $g_{n}$ and $g_{p}$ are taken to be approximately constant. It is noted that for $k_{1}=k_{3}=0$ (or $k_{2}=k_{4}=0$) we obtain the case of coupled laser systems with a single optoelectronic feedback.\\
In figure 1  a schematic diagram of the experimental set-up for synchronization is given. In the experimental scheme unidirectional coupling can be realized by inclusion of an optical isolator (OI) as shown in the figure.\\ 
First we find the existence conditions for complete synchronization between uni-directionally coupled multiple time delay lasers with optoelectronic feedbacks. Comparing Eqs. (1-2) and Eqs.(3-4) one finds that a synchronous solution
\begin{equation}
S_{2}(t) = S_{1}(t) ,     N_{2}(t) = N_{1}(t)                                                                                                               
\end{equation}
exists if
\begin{equation}
k_{1}=k_{3} + K,  k_{2}=k_{4}, \tau_{1}=\tau_{3}                                                                                                         
\end{equation}
The synchronous solution (6) also  exist if
\begin{equation}
k_{1}=k_{3},  k_{2}=k_{4} + K, \tau_{2}=\tau_{3}                                                                                                         
\end{equation}
Thus the logic cell in this paper consist of the uni-directionally coupled laser systems with two optoelectronic feedbacks. The logic output can be 
described by both the synchronization error and the cross-correlation function between the systems to be synchronized. Here in this paper the parameter of the master system, Eqs.(1-2) e.g. the bias current $I_{1}$ is determined by the input sets, and the bias current $I_{2}$ of the slave system ,Eqs.(3-4) 
acts as a logic gate controller. For this, we rewrite the bias current of the master system in the following form:
\begin{equation}
I_{1}=I_{a} + L_{1}I_{b} + L_{2}I_{c}                                                                                                        
\end{equation}
For input set (0,0) $L_{1}=0,L_{2}=0$. Input set (1,1) corresponds to the case $L_{1}=1,L_{2}=1$. Input sets (0,1)/(1,0) can be described by 
$L_{1}=0,L_{2}=1$ or by $L_{1}=1,L_{2}=0$.\\
Now suppose that as a result of the input sets (0,0), (0,1)/(1,0), and (1,1) the bias current $I_{1}$ of the master laser 
equals to reasonable separated values $I_{m1},I_{m2},$and $I_{m3}$, respectively. As emphasized above, both the synchronization error and the cross-correlation function between the synchronized lasers can be considered as the logic output of the logic cell. Let us assume that for large error the output is 0;for nearly zero error the output is 1. In other words, when synchronization between the laser occurs we obtain the output 1;otherwise one obtains output 0. Note that when the lasers are synchronized (the case of very small synchronization error), the cross-correlation function is equal to 1, otherwise (for the case of large synchronization error) it is smaller than 1.\\
For the particular logic gate the logic output will be determined by the slave(response) laser's bias current.
For example, for NOR logic gate $I_{2}$ will be set close to $I_{m1}$;for XOR gate $I_{2}$ is set close to $I_{m2}$;
with $I_{2}$ close to $I_{m3}$ we will obtain AND logic gate. 
\begin{center}
III. NUMERICAL SIMULATIONS AND DISCUSSIONS
\end{center}
In the numerical simulations we use typical values for the internal parameters of the master laser:
$\gamma_{c}=2 ps,\gamma_{s1}=2 ns, \Gamma =0.3, g_{0}=7\times 10^{13},g_{n}=10^4,g_{p}=10^4, N_{0}=1.7\times 10^{8},S_{0}=5\times 10^{6}, 
I_{1}=3.4\times 10^{17} [9].$ The parameters at the receiver are chosen to be identical to those of the transmitter, except for the feedback levels.\\
The numerical simulation results for unidirectionally coupled lasers with multiple optoelectronic feedbacks, Eqs.(1-4) 
are presented in Figures 2-3 with other parameters as $\tau_{1}=\tau_{3}=3ns, \tau_{2}=5ns, k_{1}=0.80, k_{2}=k_{4}=0.15, k_{3}=0.15, K=0.65.$
As mentioned above, our approach is based on the mismatch between the bias currents for the master and slave lasers.
In this paper we restrict ourselves to the case of complete synchronization.To characterize synchronization quality between the master and slave lasers we use two approaches; first approach is based on 
the maximum synchronization error between the intensities of lasers to be synchronized; second approach uses the cross-correlation function between the master laser and slave laser intensities.\\ 
Figure 2 shows maximum synchronization error (MSE) for the three cases of parameter mismatches. First we put the bias current of the master laser at $I_{m1}=3.4\times 10^{17}$ and calculate the MSE $e_{S}=S_{2}-S_{1}$  ($\triangle $) for the case of slave laser bias current $I_{2}$ change from  $I_{m1}$ to $I_{mf}=2.4\times I_{m1}$. Then we set the bias current of the master laser at $I_{m2}=1.7\times I_{m1}$ and again calculate maximum synchronization error $e_{S}$ ($\diamondsuit $) while changing the bias current of the slave laser in both directions  from $I_{m2}$ to $I_{f}$ and from $I_{m2}$ to $I_{m1}$. Finally, the bias current of the master laser is set to $I_{m3}=2.4\times I_{m1}=I_{mf}$ and the MSE $e_{s}$ ($\star $) is calculated for changing the bias current of slave laser from $I_{m3}$ back to $I_{m1}$. To characterise the degree of 'largeness' of the synchronization error we set the maximum synchronization error threshold at $e^{t}_{S}=0.015$, i.e. the logic output is 1 if the synchronization error is less than $e^{t}_{S}$, otherwise we obtain the logic output 0.\\
\indent Now we discuss in detail how e.g. NOR logic gate can be constructed. We have three values of the master laser's bias current reasonable separated from each other:$I_{m1},I_{m2},$ and $I_{m3}$ corresponding to input sets (0,0),(0,1)/(1,0), and (1,1), respectively. Let us take the slave laser bias current $I_{2}$ close to $I_{m1}.$ Then as follows from figure 2 (curve with $\triangle$) MSE corresponding to this case is close to zero. That is for (0,0) input we obtain near zero maximum synchronization error, which corresponds to logic output 1. If we guide our eyes 
along the ordinate axis, then we note that MSE corresponding to (0,1)/(1,0) (curve with $\diamondsuit $) input will be around 0.04, that is  well above the denoted threshold synchronization error $e^{t}_{S}=0.015$. Thus for input (0,1)/(1,0) we obtain large synchronization error, which means logic output 0. Further for input (1,1) again guiding our eyes along the ordinate axis we note that MSE is well above $e^{t}_{S}$ (curve with $\star$). It means that we again arrive at the large synchronization error case, which gives logic output 0. With all three input sets considered with the help of figure 2 we have demonstrated construction of logic gate NOR.\\
\indent Next we show how the logic gate XOR can be obtained. This time let us set the slave laser's bias current $I_{2}$ close to $I_{m2}$ (point 1.7 at the abscissa axis, figure 2). Now if we guide our eyes along the perpendicular line to the abscissa axis at point 1.7 we see that MSE corresponding to (0,0) input (curve with $\triangle$) is very large, that we obtain logic output 0. In the case of (0,1)/(1,0) input (curve with $\diamondsuit $) MSE around point 1.7 is zero, that is one obtains logic output 1. Next for input (1,1) (curve with $\star$) guiding our eyes along the vertical to the abscissa axis at 1.7 we easily establish that MSE is large, that is logic output 0 is obtained. Thus in this paragraph we demonstrated the construction of XOR gate.\\
\indent Finally we discuss AND logic gate construction. For that purpose we set the slave laser's bias current at close vicinity of $I_{m3}$, point 2.4  
at the abscissa axis (figure 2). Once more guiding the eyes along the perpendicular line to the abscissa axis at point 2.4, it is straightforward to 
establish input set (0,0) gives large MSE (curve with $\triangle$), and therefore logic output 0. Logic output 0 is also obtained for input 
set (0,1)/(1,0), as the intersection point of the perpendicular line to the abscissa axis at 2.4 with the curve with $\diamondsuit $ corresponds to large MSE.
For input (1,1) (curve with $\star$) close to slave laser's bias current around 2.4  one obtains very small (near zero) MSE, which means logic output 1.
Thus in this paragraph it is show how logic gate AND can be constructed using synchronization of optoelectronic feedback laser with parameter mismatches.\\
In figure 3, as a alternative to the use of the concept of synchronization error for construction of the logic gates 
we propose an approach based on the cross-correlation function $C$ between the laser systems to be synchronized. The procedure to construct logic gates discussed above is fully identical to that for figure 2. It is only noted that in this case logic output 1 corresponds to the value of C very close 1, and it may be a bit more straightforward approach(remind that within the synchronization error approach logic output 1 corresponds to zero (small) synchronization error). Also note that in this case the threshold value of the cross-correlation function $C$ established at the level of $C^{t}=0.9989$. That is if the value of the cross-correlation exceed $C^{t}$ then the logic output is 1, otherwise one obtains logic output 0. \\
In figure 4 (b) we portray the synchronization error dynamics for the case of input signal of rectangular pulses (figure 4(a)). Rapid changes of the synchronization error from the non-zero value (logic output 0) to near zero value (logic output 1) confirms that logic gates based on laser system studied in the paper can be quite fast (1-5 GHz) in comparison with logic gates based on Chua circuits (several kHz).\\
\begin{center}
IV.CONCLUSIONS
\end{center}
\indent In conclusion, we have demonstrated that chaos synchronization in multiple time delayed lasers with optoelectronic feedbacks can be exploited to create fast(nanoscale) dynamical flexible logic gates. Specifically we have shown that using uni-directionally coupled lasers and changing logic control parameter values of the slave laser, flexible implementation of the basic logic gates NOR, XOR, and AND is possible. Such a scheme can serve as an integral part of more flexible computing systems than statically wired hardware.\\
The results of the paper provide the basis for the use of the optoelectronic feedback lasers with multiple variable time delays in high-speed communication systems.\\
\begin{center}
V.ACKNOWLEDGEMENTS
\end{center}
This research was supported by a Marie Curie Action within the 6th European Community Framework Programme Contract.\\
\newpage
\begin{center}
Figure captions
\end{center}
TABLE 1. Relationship between the inputs and output of the NOR,XOR, and AND logic gates.\\ 
~\\
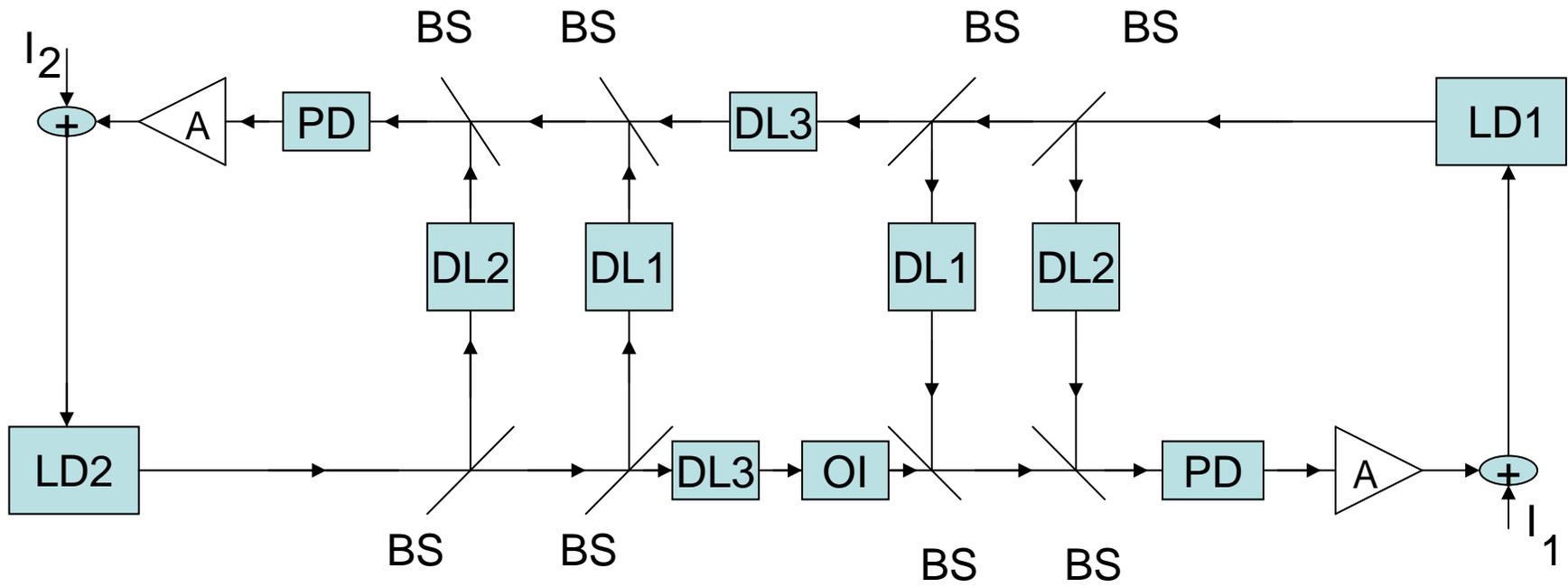
FIG.1.Schematic experimental arrangement for the synchronization of lasers with double optoelectronic feedback and injection. LD: Laser diode. PD:Photodetector.BS:Beamsplitter. DL:Delay lines. OI:Optical Isolator. A:Amplifier.$I_{1,2}$ is the bias current for the LD 1 and LD2
The output from each laser is split by a beamsplitters and directed along different feedback loops and coupling loops. Each signal is converted into  an electronic signal by a photodetector and then amplified before being added to the injection current of the laser. \\
~\\
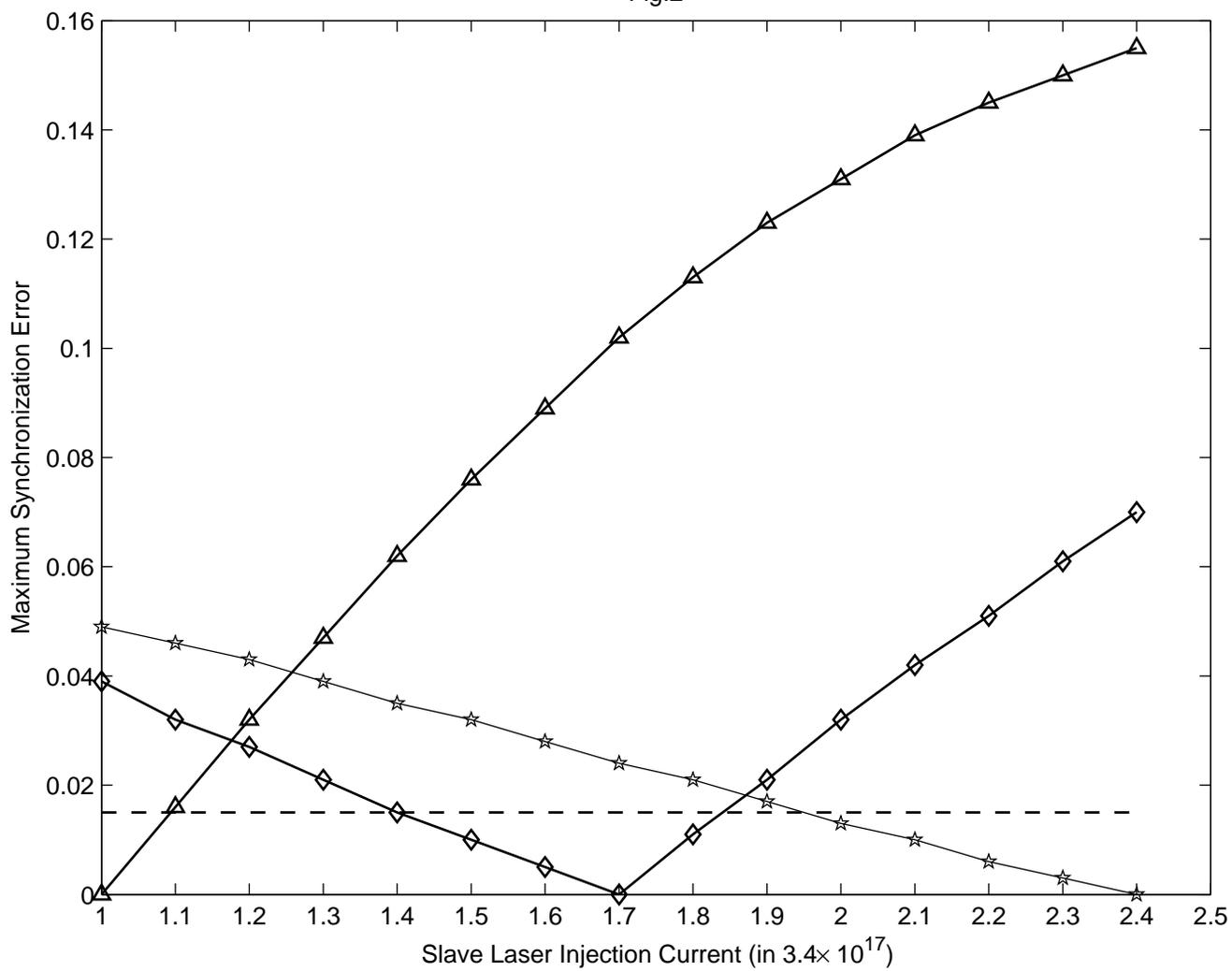
FIG.2. Numerical simulation of uni-directionally coupled  time delay lasers, Eqs.(1-4) for $\tau_{1}=\tau_{3}=3ns, \tau_{2}=5ns, k_{1}=0.80, k_{2}=k_{4}=0.15, k_{3}=0.15, K=0.65.$ Complete synchronization: Maximum synchronization error as a function of parameter mismatches between the bias currents for the master and slave lasers (see text for details)\\
~\\
FIG.3. Numerical simulation of uni-directionally coupled  time delay lasers, Eqs.(1-4) for $\tau_{1}=\tau_{3}=3ns, \tau_{2}=5ns, k_{1}=0.80, k_{2}=k_{4}=0.15, k_{3}=0.15, K=0.65.$ Complete synchronization: The cross correlation function vs parameter mismatches between the bias currents for the master and slave lasers (see text for more details)\\
~\\
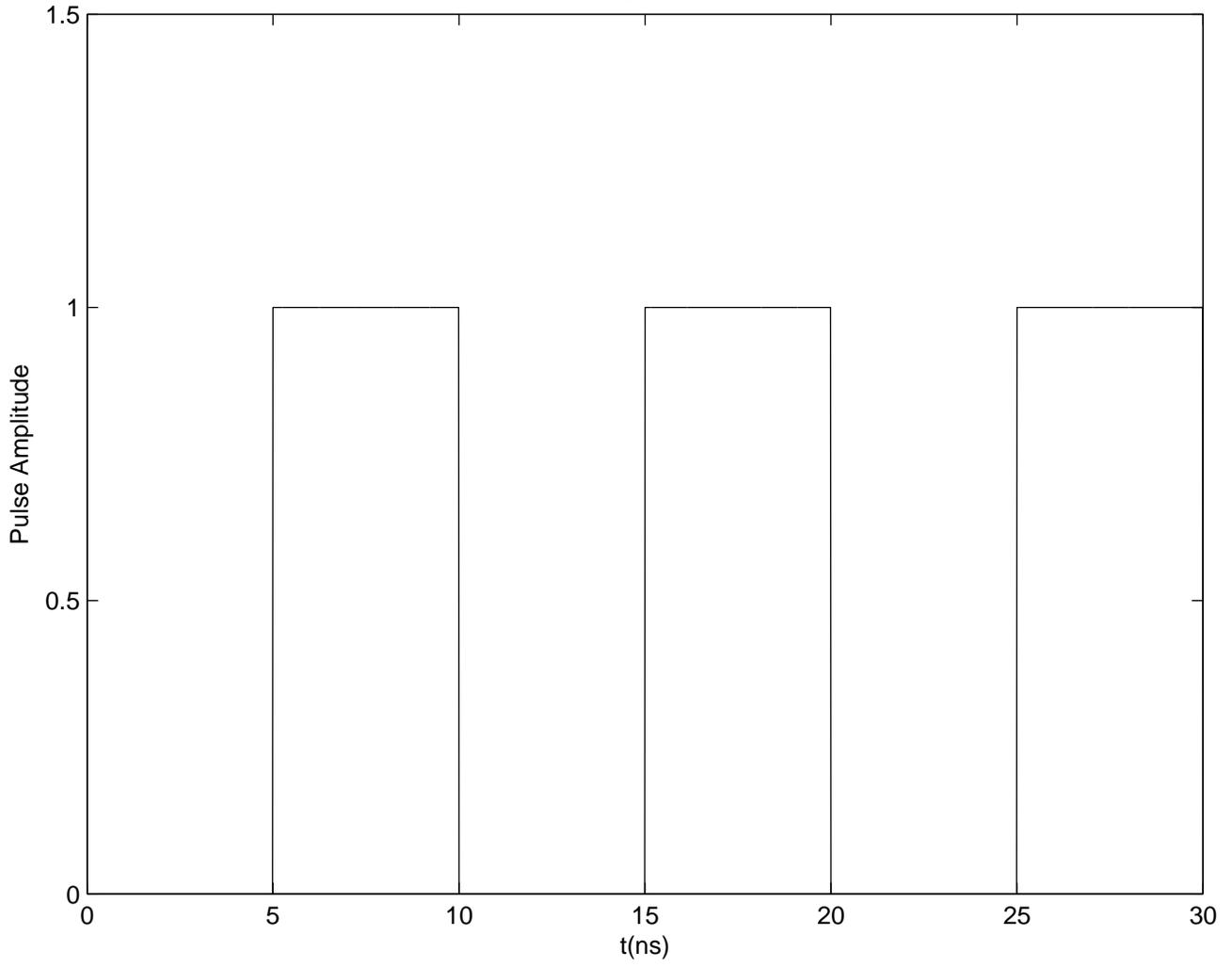
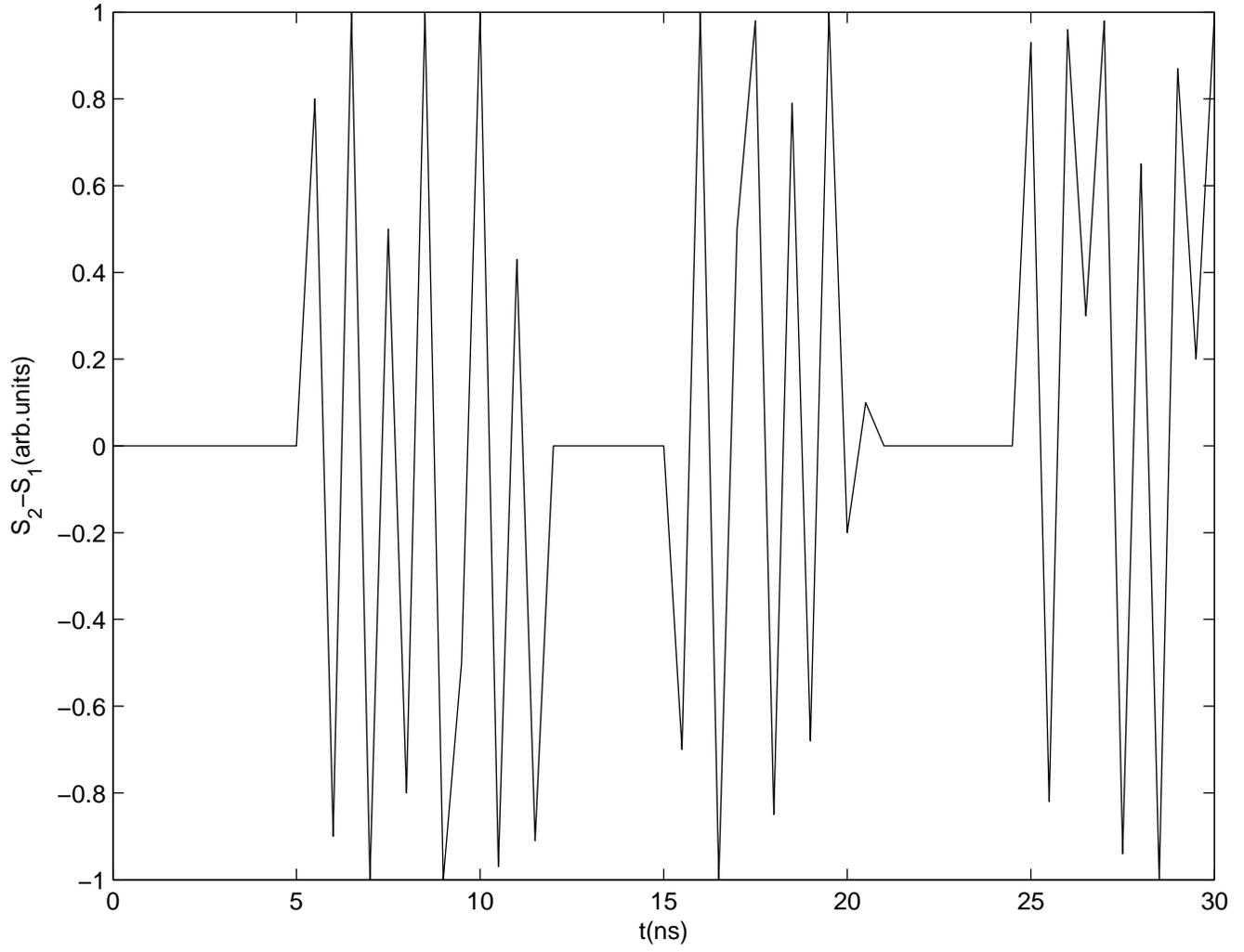
FIG.4. Numerical simulation of uni-directionally coupled  time delay lasers, Eqs.(1-4) for $\tau_{1}=\tau_{3}=3ns, \tau_{2}=5ns, k_{1}=0.80, k_{2}=k_{4}=0.15, k_{3}=0.15, K=0.65.$ Complete synchronization: Synchronization error dynamics (fig.4(b)) for the case of input signal of rectangular pulses (fig.4(a)) .\\

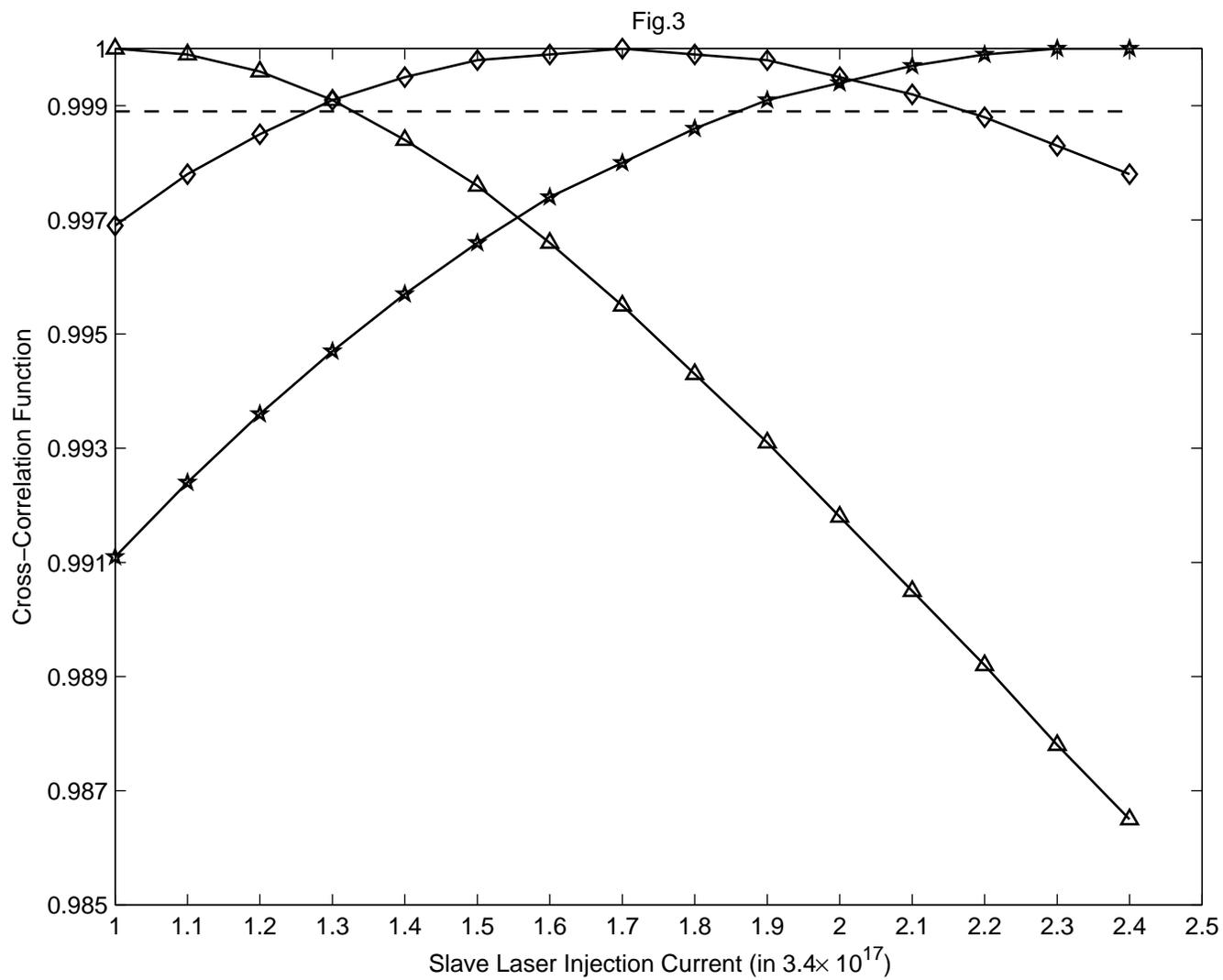
\end{document}